# Very Low Mass Stars: From Observations to Theory and back

By E. BROCATO[1,2], S. CASSISI[1,3], AND V. CASTELLANI[1,2,4]

[1]Osservatorio Astronomico Collurania, Via M. Maggini, 64100 Teramo, Italy

[2]Istituto Nazionale di Fisica Nucleare, LNGS, 67100 L'Aquila, Italy

[3]Dipartimento di Fisica, Universitá de L'Aquila, Via Vetoio, 67100 L'Aquila, Italy

[4]Dipartimento di Fisica, Universitá di Pisa, Piazza Torricelli, 56100 Pisa, Italy

We briefly review the observational and theoretical investigations on Very Low Mass (VLM) stars. New evolutionary computations, extending toward the lower main sequence previous computations already given for globular cluster stars, are presented. Comparisons with updated observational data both for globular cluster stars and VLM stars with known parallaxes are also shown. We discuss observational evidences supporting theoretical predictions concerning the Color - Magnitude (C-M) location of faint Main Sequence stars as far as concern the dependence of the lower main sequence on the metallicity.

## 1. Introduction

As it has been early recognized, the largest amount of stars in our Galaxy is formed by low mass stars still burning H at the center of their rather unevolved main sequence structures. However, investigating the lower main sequence has been a tantalizing goal for a long time. As a matter of the fact, for a long time observations have been hampered because the intrinsic faintness of these structures. In the mean time, it was early understood that similar simple stellar structures do require a particular sophisticated physics to describe the state of the gas at low temperatures and relatively high densities as well as the radiative opacity affecting their cool atmospheres.

The first light on that problem came from observation. After some pioneering works (see, e.g., Spinrad 1973, Upgren & Weis 1975, Veeder 1974, Mould & Hyland 1976) we were progressively facing an increasing body of observational data, as finally given by the rather large sample of stars with known parallaxes (Monet et al. 1992, Dahn et al. 1995). This sample already disclosed the occurrence of a well populated and, rather well defined sequence of stars to be interpreted as solar metallicity Low Main Sequence (LMS) objects, with the additional evidence for metal poor subdwarfs spreaded toward higher effective temperatures. Such an evidence obviously stimulated the theoretical competition to fit the data. The first attempts (Grossman 1970, Hoxie 1970, Copeland et al. 1970, Grossman & Graboske 1971, Hoxie 1973, Grossman et al. 1974) were revisited and largely improved (VandenBerg et al. 1983, D'Antona 1987, Dorman et al. 1989, Burrows et al. 1993, Nelson et al. 1993) to address the problem of deriving as more realible as possible theoretical models of VLM stars. Although the agreement with the observations improved significantly within the past decade or so, a discrepancy still exists in all these models, which usually predict too large effective temperatures for a given luminosity.

Whereas theory is with some difficulty reaching observations, following the improved efficiency of the Hubble Space Telescope one is now facing a tremendous improvement in the observation of stars at the faint end of the globular cluster main sequence. The recent HST observations (NGC6397 - Paresce et al. (1994) & Cool et al. (1996), M30 - Piotto & King 1996) have disclosed a well defined and well populated sequence which should put tight observational constraints on the evolutionary behavior of metal poor, low main sequence (LMS) stars. Thus again observation is stimulating new theoretical work.

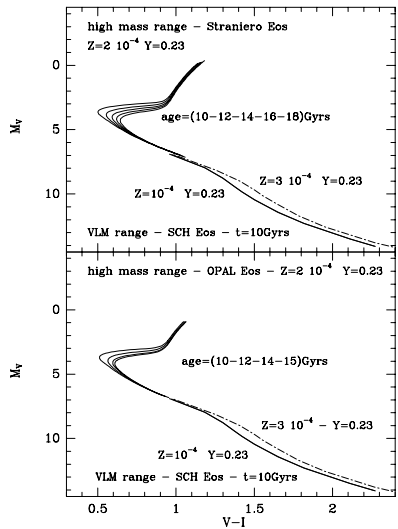

FIGURE 1. Comparison between isochrones obtained using the OPAL Eos and the SCH Eos (*Bottom panel*) and, isochrones computed adopting the Straniero Eos and the one of SCH (*Top panel*).

However, the suggestion for revisiting the theoretical evolutionary scenario for VLM stars follows the evidence for recent and significant improvements in the physical inputs or, more precisely, in both the opacity and the EOS. As for the EOS, Saumon *et al.* (1995 hereinafter SCH) have recently provided an accurate EOS appropriate for dense and cool stellar objects, which should represent a relevant improvement in comparison with other EOSs suitable for dense and cool objects. It is interesting to notice that SCH EOS provides, when used in computing stellar models, results in good agreement with the ones obtained adopting the EOS by Rogers *et al.* (1996, OPAL EOS) for computing more massive stellar models. In figure 1, we compare isochrones obtained from stellar models computed adopting the SCH EOS in the VLM stars' range and alternatively the new OPAL EOS or the 'old' Straniero (1988) EOS for the models with mass equal or greater than $0.6 M_\odot$. The good match obtained between the OPAL EOS models and the SCH EOS ones largely improves previous evaluations.

As far as opacity is concerned, Alexander (1994) and Alexander & Ferguson (1994) computed Rosseland mean opacities, which account accurately for both atomic and molecular absorbers. Relying on such 'fresh' physical inputs, it is obviously interesting to explore whether or not the current evolutionary scenario is changed and it provides a better agreement with the new observational data. However, it is worth noting that another further complication in computing models of VLM objects is related to the treatment of the atmosphere (Burrows *et al.* 1993, Dorman *et al.* 1989, Baraffe *et al.* 1995). In the atmosphere of these cool stars one expects that convection can deeply affects atmospheric layers for two reason: i) the adiabatic gradients are small due to the dissociation of $H_2$ and, ii) the opacity can became large due to the presence of molecules. Therefore, in structures less massive than about $0.5 M_\odot$ the atmosphere can become unstable against convection at very low optical depth ($\tau \approx 0.01$) and the use of normal $T(\tau)$ relation is not more valid. The solution of this problem relies on the possibility of incorporating model atmospheres (Allard & Hauschildt 1995, Brett 1995, Brett & Plez 1993) in stellar evolutionary codes (see Baraffe *et al.* 1995 and Chabrier *et al.* 1996). However, significative discrepancies do still exist between different sets of model atmosphere (Bessel 1995) due mainly to the different techniques adopted in computing the opacity. So we choose for the moment to keep on the Eddington approximation, i.e. to adopt a $T(\tau)$ relation (in particular the Krishna Swamy (1966) one) which describes the temperature stratification in the outermost layers. As it will be shown in the following, we think that this assumption has to be regarded as a first order, of course, but not to bad approximation.

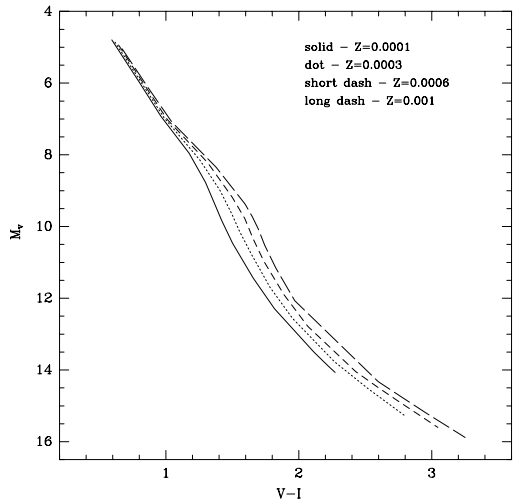

FIGURE 2. Theoretical expectations for the (V,V-I) color - magnitude diagrams of faint MS stars for the labeled assumtions on the metallicity.

## 2. Theoretical VLM stellar structures.

Selected sequences of stellar models have been computed to cover the range of hydrogen burning stars below $0.8 M_\odot$. Computations have been performed adopting the following metallicities: Z=0.0001 - 0.0003 - 0.0006 - 0.001 - 0.006, and a Helium abundance $(Y)$ equal to 0.23 everywhere. A sequence of models with solar metallicity and Y=0.28 has been also computed. According to the calibration of solar models (Salaris & Cassisi 1996), in all computations it has been adopted a mixing length parameter as given by $\alpha = 2.2$. The models have been transferred from the theoretical plane to the observational one by adopting bolometric corrections and color-temperature relation by Allard & Hauschildt (1995) and by Kurucz (1992). Figure 2 reports the location of 10 Gyrs sequences in the observational $(M_V, V-I)$ plane for the labelled assumptions on the metallicity. Further details on these models, as well as a detailed description of VLM evolutionary features can be found in Alexander *et al.* (1996). However in the following section we wish to address the attention of the reader toward two unexpected evolutionary properties of VLM stellar models.

### 2.1. *Some interesting features....*

As it is already known, during the H burning phase stars less massive than about $0.3 M_\odot$ are fully convective. Analizing in detail the behavior of similar stellar structures one finds a peculiar behaviour of the density against the time. As a matter of fact, in "normal" models (more massive than $0.3 M_\odot$) the consumption of central H drives a continuos increase of both central temperature and central density. However, in fully convective structures the central density keeps decreasing all along the major phase of H burning. It has been found (Alexander et al 1996) that such 'anomalous' behavior is related to the increase of the molecular weight in the whole structure due to both the convective mixing and the production of Helium by nuclear reactions. It may be worth noting that such behavior of the central density is quite similar to the one of central density during the first phases of He burning in more massive stars.

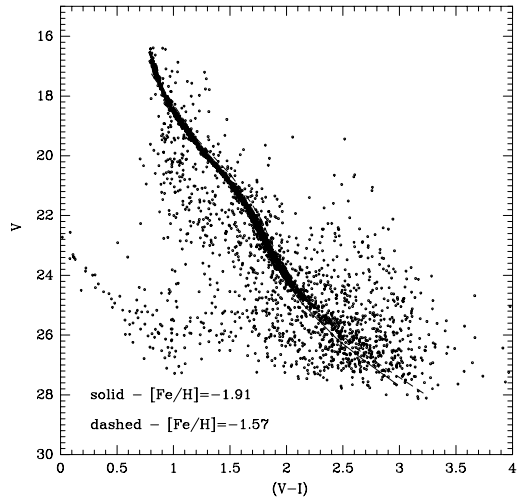

FIGURE 3. (V, V-I) CM diagram for lower main sequence stars in the globular cluster NGC6397. The comparison with theoretical sequences for t=10 Gyr and for the two labeled assumptions about the star metallicity has been performed adopting for the cluster $(m - M)_V = 12.4$, $E(V - I) = 0.19$.

By discussing the effect of the age on the HR diagram location of VLM stars, one finds the well known feature that age plays a negligible role for structures older than about 10Gyrs. But one can notice the further curious property for which the $0.3 M_\odot$ model appears as the less affected by the variation of the age, while less massive models show a progressively increasing sensitivity to the age. One understands such an occurrence when accounting for the role played by the time evolution of the $^3He$ abundance in these stars (for more details see Alexander *et al.* 1996).

## 3. Comparison with the observational data.

### 3.1. *NGC6397 and metal poor VLM models.*

In figure 3, we show the comparison between the HST data for NGC6397 by Cool *et al.* (1996) and our theoretical prescriptions for two selected metallicity. The fine agreement obtained between the observational data and the theoretical sequence for $[Fe/H] = -1.91$ appears largely satisfactory. Note that the agreement has been achieved for values of both reddening and distance modulus well compatible with the current estimates.

### 3.2. *VLM with known parallaxes and solar VLM models.*

The sample of stars with known parallaxes has been for a long time a fundamental tool to check the reability of the theoretical scenario for VLM stellar objects. In figure 4, we show the most complete, presently available, color - magnitude diagram for such stars from the extensive parallax survey being carried out by the US Naval Observatory (Monet et al. 1992, Dahn et al. 1995). In the same figure, we report also our theoretical predictions for solar composition VLM stars together with similar data as presented by by D'Antona & Mazzitelli (1994, 1996) and by Baraffe et al. (1995). These last Authors use the same EOS and opacity evaluations as in our computations, but adopt model atmosphere as outer condition.

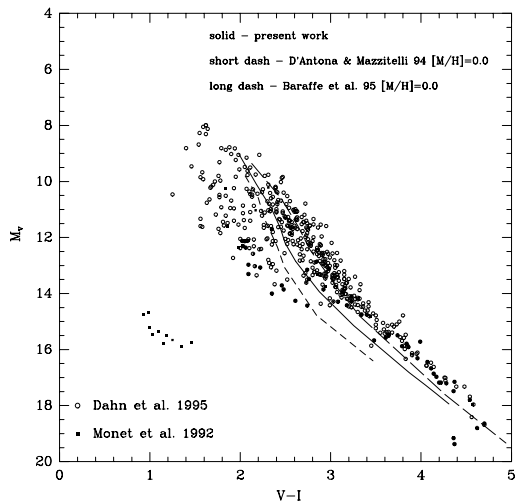

FIGURE 4. ($M_V$,V-I) CM diagram for faint stars with known parallaxes from Monet et al. (1992) and Dahn et al. (1995) with superimposed theoretical distributions for solar metallicity models.

All models fail in accurate reproducing the location of the lowest MS masses. This occurrence is probably due to the difficult task of taking into the right account the proper influence of metals on both the opacity and the EOS. This occurrence suggests that the knowledge of the evolutionary scenario for VLM stars is not yet complete, so remaining a tantalizing goal for stellar theoricists.

## 4. The ranking of low luminosity Main Sequence stars with the metallicity.

As it is well known, a fundamental feature of the Red Giant Branch (RGB) in Galactic globular is its sensitive dependence on the metallicity of the stars in the cluster: the cluster RGBs become bluer and steeper when the metallicity of the cluster decreases. However, all recent theoretical works on VLM stars (Baraffe et al. 1995, D'Antona & Mazzitelli 1996, Alexander et al. 1996) have shown that the location in the H-R diagram of such stars is also expected to depend sensitively on the metallicity, again becoming bluer and steeper when the metallicity is decreased (see figure 2).

In the following, we will report the evidence that such a prediction on the Low Luminosity Main Sequences is eventually supported by recent accurate photometries of galactic globular cluster (for more details see Brocato et al. 1996). We have selected a little sample of clusters with accurate photometries for the Main Sequence, namely: NGC6397 (Cool et al. 1996), M30 (Piotto & King 1996), M4 and M55 (Fahlman et al. 1996). In table 1 we report some important parameters for these clusters. Metallicities have been taken from the compilation provided by Carretta & Gratton (1996) for all cluster, but M55, for which we adopt the value reported in the work by Chaboyer & Kim (1995). As far as the distance modulus and reddening, we adopt the current values in the literature. According to the data in table 1 and accounting for a possible indetermination on the metallicity of about 0.10dex, the clusters NGC6397, M55 and M30 may well have the same metallicity. As a matter of the fact, by overplotting the C-M diagrams of these clusters one finds a nice global ovelap between the three stellar distributions, once the differential differences in both reddenings and distance moduli have been accounted for.

| Cluster | [Fe/H] | $(m-M)_V$ | E(B-V) |
|---------|--------|-----------|--------|
| M4      | -1.19  | 12.65     | 0.40   |
| NGC6397 | -1.82  | 12.50     | 0.16   |
| M55     | -1.82  | 14.10     | 0.14   |
| M30     | -1.91  | 14.85     | 0.01   |

TABLE 1. Values of the metallicity, distance modulus and reddening for the selected globulars

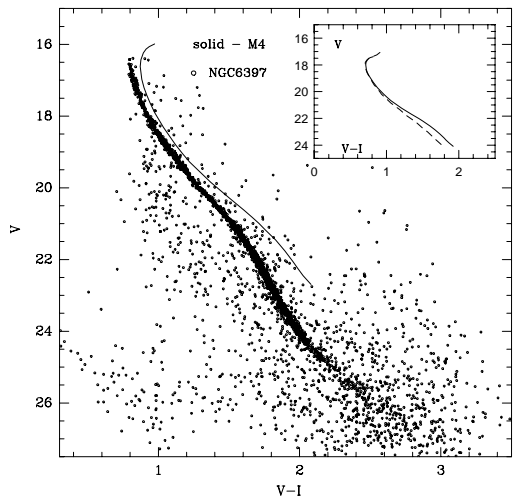

FIGURE 5. Comparison between the C-M diagrams of NGC6397 and M4 with M4 reduced to the same reddening and distance modulus of NGC6397. In the box, we present the comparison between the mean lines of M4 and M55 (see text for more details).

As a most interesting point, let us compare the metal poor cluster NGC6397 with the more metal rich M4. Figure 5 shows that at the fainter luminosities the M4 Main Sequence clearly develops toward much redder color than NGC6397. The amount of such effect is so large that it can not be related to observational errors and/or indeterminations in the trasformation of the HST colors. To go deeper in this investigation, in the same figure we compare the two clusters M4 and M55 simply applying a shift to the stellar distributions for obtaining a close overlap in the Turn Off region. Once again we find that the MS slope of M4 is lower than the M55 one.

So, as expected on theoretical grounds, we conclude that the Low MS of globular clusters is a sensitive indicator of the cluster metallicity as the Red Giant Branch.